\documentclass[11pt]{article}
\usepackage{jheppub}
\usepackage{hyperref}
\usepackage{amsmath}
\usepackage{amsfonts}
\usepackage{amssymb}
\usepackage{mathrsfs}

\usepackage{package-notes}

\hypersetup{linkcolor=red,anchorcolor=green,citecolor=blue,urlcolor=blue}

\def\d{\partial}
\newcommand{\im}{\mathrm{i}}
\newcommand{\sigp}{\sigma^{+}}
\newcommand{\sigm}{\sigma^{-}}
\def\bms3{\ensuremath{\mathfrak{bms}_3}}
\def\gca2{\ensuremath{\mathfrak{gca}_2}}
\newcommand{\mL}{\mathcal{L}}
\newcommand{\vacp}{|0\rangle_{HS}}
\newcommand{\vacxp}{|0\rangle_{{A}}}

\begin{document}

\title{On the null origin of the ambitwistor string}
\author{Eduardo Casali,${}^1$Piotr Tourkine${}^2$}
\vspace{.4cm}
\affiliation{$^1$Mathematical Institute, University of Oxford, Woodstock Road, Oxford OX2 6GG, UK\\
${}^2$Department of Applied Mathematics and Theoretical Physics,
Wilberforce Road, Cambridge CB3~0WA,
UK.}
\emailAdd{casali@maths.ox.ac.uk}
\emailAdd{pt373@cam.ac.uk}
\preprint{DAMTP-2016-45}

\abstract{In this paper we present the null string origin of the
  ambitwistor string. Classically, the null string is the tensionless
  limit of string theory, and so too is the Ambitwistor string. Both
  have as constraint algebra the Galilean Conformal Algebra in two
  dimensions. But something interesting happens in the quantum theory
  since there is an ambiguity in quantizing the null string. We show
  that, given a particular choice of quantization scheme and a
  particular gauge, the null string coincides with the ambitwistor
  string both classically and quantum mechanically. We also show that
  the same holds for the spinning versions of the null string and
  Ambitwistor string. With these results we clarify the relationship
  between the Ambitwistor string, the null string, the usual string
  and the Hohm-Siegel-Zwiebach theory.}
\maketitle

\section{Introduction}
\label{sec:introduction}

The Cachazo-He-Yuan (CHY)
formulae~\cite{Cachazo:2013iea,Cachazo:2013hca,Cachazo:2013gna,Cachazo:2014xea,Cachazo:2014nsa}
are a remarkable set of formulae, expressing the tree-level
scattering amplitudes of a variety of theories (scalar, gauge,
gravity) as worldsheet integrals localized onto the solutions of the
so-called scattering equations. The Mason-Skinner Ambitwistor string
\cite{Mason:2013sva,Berkovits:2013xba} gave a rationale for these formulae as a
holomorphic string subjected to the constraint $P(z)^2=0$, which
enforces the scattering equations.\footnote{$P^\mu(z)$ is the momentum of the string.} This stringy origin allowed the CHY formalism to be extended
in several directions \cite{Casali:2015vta,Adamo:2014wea,Adamo:2015hoa}, including, very interestingly, loops \cite{Adamo:2013tsa,Casali:2014hfa,Geyer:2015bja,Geyer:2015jch}. 

So far, all the theories describable in this
elegant formalism are, more or less, compactifications of type~II
supergravity or super-Yang-Mills. It is therefore natural to expect
that Ambitwistor string could correspond to some field theory limit,
i.~e. $\alpha'\to0$, of string theory.

However, the scattering equations are also famously known to govern
the dynamics of the opposite limit of string theory, the Gross-Mende
limit, where the tension $T=1/(2\pi\alpha')$ goes to
zero~\cite{Amati:1987wq,Gross:1987ar}. How then could these two limits
have anything in common ?

In this paper, we argue that this paradox is clarified when the
Ambitwistor string is seen as a tensionless string. Already long
ago~\cite{Gamboa:1989zc,Gamboa:1989px} it was noticed that null
strings~\cite{Schild:1976vq}, or tensionless strings, possess a
quantization ambiguity that can lead to two very different theories.
The first one is a
higher-spin like theory~\cite{Lizzi:1986nv}, in accordance with the picture of
\cite{Gross:1988ue}. And a truncated theory with only a few states in
its spectrum. In the case of the RNS -- or spinning -- null string, this
spectrum was observed to be that of type II supergravity
in~\cite{Gamboa:1989zc}. We review these constructions in this paper
and relate them to the Ambitwistor string and the scattering
equations.

In the tensionless limit, the 
classical tensionful string constraint $P^2+T^2 (\partial_\sigma X)^2=0$,
reduces to $P^2=0$, therefore it is natural that both the
Gross-Mende limit and the Ambitwistor string have the scattering
equations at the heart of their dynamics. This happens to be so, even
though they are different quantum theories, with
Fock-spaces built out from different vacua. In order for the reader to
appreciate this point already now, we give below the definitions of
the two different vacua of the quantized tensionless string in terms
of the Fourier modes of the $P^\mu$ and $X^\mu$ fields;
\begin{equation}
  \label{eq:vacua}
  p_n^\mu \vacp=0\,,  \quad\forall n\in\mathbb{Z}\qquad \mathrm{\mathit{vs}} \qquad
  \begin{cases} x_n^\mu\vacxp=0\\ p_n^\mu \vacxp=0
    \end{cases}\quad
    \forall n>0\,,
\end{equation}
where the indices $HS$ and ${A}$ stand for ``higher-spin like''
and ``Ambitwistor'', respectively.%
A significant part of this paper is concerned with a study of the
various relationship between these theories, at the classical and
quantum level. 

Interestingly, the constraint
algebra of the tensionless string has been studied recently in
particular in the context of the flat holography, and is
nowadays known as the Galilean Conformal Algebra~\cite{Bagchi:2009my,Bagchi:2013bga}.
This algebra usually appears from the non-relativistic limit of a
conformal field theory, but there exists an interesting coincidence in two dimensions.
Two different contractions of the Virasoro algebra, a non-relativistic one where the worldsheet speed of light
goes to infinity, and a ultra-relativistic one where it goes to zero, give isomorphic algebras. 
One could therefore expect that these two limits should, naively,
correspond to the two different tensionless theories. But this is not necessarily the case. In fact, both limits give locally gauge-equivalent
versions of the same theory, and moreover, the Ambitwistor string is
yet another gauge choice for such system. We argue that this gauge
choice is an important element as it allows to identify one of the
null string constraint with the holomorphic stress tensor of a chiral
conformal field theory.

The Ambitwistor string is therefore not the limit of the usual string
at the quantum level, but rather the limit of an alternatively
quantized string as will be expanded below. This quantization scheme
was also found by Siegel in \cite{Siegel:2015axg}, where it was
referred to as a change of boundary conditions.

In the following sections we review the classical and quantum aspects of
the tensionless or null string, based on older results in the
literature. We then reinterpret the Ambitwistor string as a
gauge-choice of the null string, clarifying some of its geometrical
aspects regarding the interplay between its intrinsically degenerate
metric and emergent light-cone structure. The relationship with the
usual tensionful string is tackled next, where we relate it to a
twisted quantization of the string and show how the normal ordering
constant is affected by such quantization scheme.

One of the most interesting applications of our work is that we expect
that it should pave the way to understanding completely the moduli
problem of the ambitwistor string, following for instance
\cite{Sundborg:1994py}. In particular, this should allow a geometrical
determination of the integration cycle at
loop-level~\cite{Ohmori:2015sha}, something that cannot be done from
purely CFT considerations. This would also clarify the question of
modular invariance in these theories.

\begin{figure}
\centering
\hspace{50pt}\includegraphics{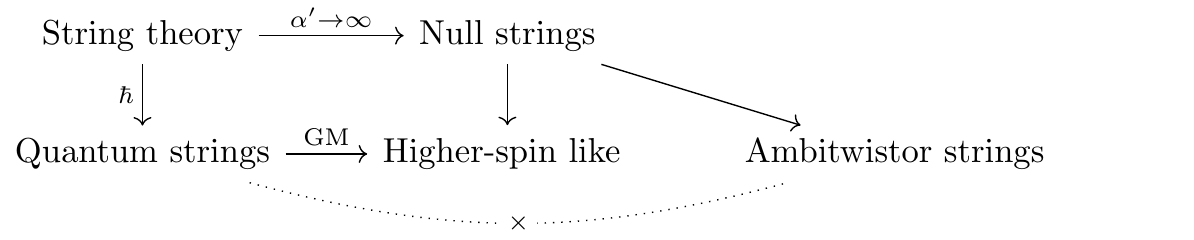}
\caption{Diagram that illustrates the ambiguity of the order between
  the limit $\alpha'\to\infty$ or Gross-Mende limit and quantization.}
\end{figure}

\section{The tensionless limit}
\label{sec:tensionless}

The tensionless limit of the bosonic string has been studied in the
past from a variety of points of view. Here we summarize the parts of
the literature relevant to the ambitwistor string. 

\subsection{Classical action}

We start with the Nambu-Goto action
  \begin{equation}
    \label{eq:nambu}
    S_{NG}=-T \int_M  d ^2\xi \sqrt{-g}\,,
  \end{equation}
  where $T$ is the tension,
  $\xi^\alpha=(\tau,\sigma)$ for $\alpha=0,1$ are the coordinates on the
  worldsheet $M$, and $g=\det g_{\alpha\beta}$ where
\begin{equation}
  \label{eq:Gamma}
  g_{\alpha\beta}=\frac{\partial X^\mu}{\partial \xi^\alpha}\frac{\partial X^\nu}{\partial \xi^\beta}\eta_{\mu\nu}
\end{equation}
is the pull-back of the space-time metric to the worldsheet. We take
the target space metric $\eta_{\mu\nu}=\mathrm{Diag}(-,+,\ldots,+)$ to
be the usual flat metric on Minkowski space. We use standard
periodicity $\sigma+2\pi\simeq \sigma$, so the worldsheet is a
cylinder. In this paper we only deal with the closed string. There are
tensionless versions of the open string, see \cite{Isberg:1993av} in particular, but
their geometry is not fully understood, and we leave their connection
to a potential open Ambitwistor string to future study.

In this action, taking $T=0$ directly is not meaningful. So, just like
in the relativistic particle case, one takes a detour and either
introduces auxiliary fields, as in
\cite{Karlhede:1986wb,Gamboa:1989zc}, or goes to the Hamiltonian formulation.
In both approaches, the tension $T$ appears linearly in the subsequent action and the limit is trivial. Here, we follow the latter, as in~\cite{Lindstrom:1990qb}.
From the action (\ref{eq:nambu}) we obtain canonical momenta
\begin{equation}
  \label{eq:P-can}
  P_\mu= T \sqrt{-g}g^{\alpha0}\partial_\alpha X_\mu
\end{equation}
that obey the following first class constraints
\begin{align}
  \label{eq:constraints}
  P^2+T^2gg^{00}=0\,,\\
P\cdot X' =0\,.
\end{align}
We use the standard notation $\partial_\tau X\equiv \dot X$ and
$\partial_\sigma X\equiv X'$.

The theory given by the Nambu-Goto action (\ref{eq:nambu}) has reparametrization invariance, so the canonical Hamiltonian vanishes. Time evolution is completely governed by the constraints, and the Hamiltonian is given by
\begin{equation}
  \label{eq:hamilt}
  \mathcal{H}=\lambda(P^2+T^2gg^{00})+\mu(P\cdot X' ),
\end{equation}
where $\lambda$ and $\mu$ are Lagrange multiplier fields. Their
arbitrariness reflects the gauge freedom of the system: setting them
to particular values amounts to choosing a particular coordinate
system on the worldsheet.  The phase-space or Hamiltonian form of the action reads
\begin{equation}
  \label{eq:S-h}
  S_H=\int d^2\xi \left(P \dot X - \mathcal{H}\right)=\int d^2\xi \left(P\cdot(\dot X-\mu
  X')-\lambda(P^2+T^2gg^{00})\right)
\end{equation}
Integrating out $P$ gives the second order action
\begin{equation}
  S=\frac12\int d^2\xi \frac1{2\lambda}\left((\dot X-\mu X')^2-4\lambda^2 T^2gg^{00}
  \right)
\end{equation}
which can be rewritten in the Polyakov form
 \begin{equation}
   \label{eq:S-P}
 S_P=-\frac{T}2\int d^2\xi \sqrt{-h}h^{\alpha\beta}\partial_\alpha
  X\cdot \partial_\beta X
\end{equation}
with the auxiliary metric
\begin{equation}
  \label{eq:polyakov-gauge}
  h^{\alpha\beta}=\begin{pmatrix}
    -1&\mu\\\mu& -\mu^2 +4\lambda^2T^2
  \end{pmatrix}
\end{equation}
The choice $\lambda=\frac1{2T},\,\mu=0$ is the conformal gauge.
Upon setting $T=0$, the metric degenerates. It is then possible to rewrite the action (\ref{eq:polyakov-gauge}) in terms of an auxiliary field $V^\alpha$, defined by
\begin{equation}
V^\alpha=\frac1{2\sqrt{\lambda}}(1,-\mu)\label{eq:V-def}.
\end{equation}
We then obtain the
Lindstr\"om-Sundborg-Theodoris~\cite{Lindstrom:1990qb} (LST) tensionless string action
\begin{equation}
  \label{eq:S-V}
  S=\int d^{2}\xi V^\alpha V^\beta \partial_\alpha X \partial_\beta X.
\end{equation}
The equations of motion (EOM) for $V^\alpha$ imply that
the metric $g_{\alpha\beta}$ has a null eigenvector. Therefore
$\det{g}=0$ and the tensionless string is a null string. Note that
$V^\alpha$ transforms as a vector density under worldsheet
diffeomorphisms, $\xi^\alpha\to\xi^\alpha+\epsilon^\alpha$;
\begin{equation}
  \label{eq:deltaV}
  \delta V^\alpha=-V\cdot \partial\epsilon^\alpha +
  \epsilon\cdot \partial V^\alpha+\frac12(\partial\cdot\epsilon)V^\alpha.
\end{equation}
This transformations maintains the diffeomorphism invariance of the
original string, though not its Weyl invariance. Although the metric
degenerate in the null string, it is still possible to obtain its stress energy
tensor using the Noether procedure and a local translation
transformation. This gives;
\begin{equation}
 \label{eq:stress_energy}
 T_{\alpha}{}^\beta=V^\beta V^{\gamma}\partial_\gamma X\cdot\partial_\alpha X -\frac{1}{2}\delta_\alpha{}^\beta V^{\gamma}V^{\delta}\partial_\gamma X\cdot\partial_\delta X\,.
\end{equation}
By construction the components of the stress energy tensor are the
$T\rightarrow0$ limit of the constraints~\eqref{eq:constraints}. So
there is no change in the fact that gravity is being gauged in the
worldsheet, but in the absence of a background non-degenerate
metric,\footnote{Or equivalently a background complex structure} the
moduli problem becomes more complicated.

In most of the literature, the gauge is fixed to
\begin{equation}
  \label{eq:V-transverse}
  V^\alpha=(1,0)
\end{equation}
which is called Schild or transverse gauge.
The question of weather these gauges can be achieved globally or not has not been studied in this context. 
However, we know from the Ambitwistor string quantization that the
obstruction to gauge fixing precisely furnish the moduli of the
problem. In \cite{Sundborg:1994py} it is was argued that the vector
fields \eqref{eq:V-def} have no moduli. While the Ambitwistor does
have moduli, they are completely fixed by the scattering equations and
the resolution of the apparent conflict might lie on this point. We will address the question of moduli for the null string in future work.

\subsection{Constraint algebra}
We review here the construction of~\cite{Lizzi:1986nv,Gamboa:1989zc}.
Imposing the transverse gauge yields the action
\begin{equation}
 \label{eq:Schild-ac}
S=\frac{1}{2}\int \dot{X}^2,
\end{equation}
which is supplemented by the first-class constraints
\begin{equation}
 \dot{X}^2=\dot{X}\cdot X'=0.
\end{equation}
The EOM read
\begin{equation}
 \label{eq:Schild-EOM}
 \ddot{X}=0\,,
\end{equation}
Together with the constraints we see that this action describes a
collection of massless particles moving at the speed of light, bound
together by a constraint forcing their velocities to be orthogonal to
the string. This last constraint is a remnant of the stringy character
of the original tensionful string.
The EOM are solved by\footnote{We omit the Lorentz index where they can be understood from context.}
\begin{equation}
 X(\tau,\sigma) = Y(\sigma) + \tau P(\sigma).
\end{equation}
The mode expansion for each field is defined as
\begin{equation}
 \label{eq:mode-exp}
 Y(\sigma) = \frac1{\sqrt{2\pi}}\sum_{n\in \mathbb{Z}}y_n e^{\im n\sigma}\,,\qquad 
 P(\sigma) = \frac1{\sqrt{2\pi}}\sum_{n\in \mathbb{Z}}p_n e^{\im n\sigma} 
\end{equation}
with Poisson brackets
\begin{equation}
\label{eq:PB}
 \{y_n,p_m\}_{PB}=\delta_{n+m,0}.
\end{equation}
The zero modes $x_0$ and $p_0$ are the centre of mass and momentum of the string. In terms of these the constraints modes read
\begin{equation}
  \label{eq:cstr-trans-gauge}
  \begin{aligned}
  \dot X^2 &=
  \frac1{2\pi}\sum_{n}\left(\sum_mp_{n+m}p_{-m}\right)e^{in\sigma}=
  -\frac1{\pi}\sum_{n}M_ne^{in\sigma}\\
  \dot X\cdot X'&=\frac1{2\pi} \sum_n \left( \sum_m p_{n+m}(-\im m
    y_{-m}-\im m\tau p_{-m})   \right)e^{in\sigma}\\
  &= \frac{1}{2\pi} \sum_n (L_n- in\tau M_n)e^{in\sigma}=0
\end{aligned}
\end{equation}
where we have defined the following classical modes for the constraints;
\begin{equation}
 \label{eq:modes-cons}
 L_l=-\im\sum_{n}(l-n)p_n\cdot y_{l-n}\,,\qquad
M_l=-\frac12\sum_{n}p_n\cdot p_{l-n}
\end{equation} 
These generate the constraint algebra
\begin{align}
 \label{eq:algebra}
\{L_k,L_l\}_{PB}=-\im(k-l)L_{k+l}\,,\quad
\{L_k,M_l\}_{PB}=-\im(l-k)M_{k+l}\,,\quad
\{M_k,M_l\}_{PB}=0\,,
\end{align}
which is nowadays known as the 2d Galilean Conformal Algebra (GCA)
\cite{Bagchi:2009my,Bagchi:2009pe}. This algebra can be obtained starting from the
two copies of the Virasoro algebra and taking a ultra-relativistic
contraction of their generators. In this limit, the small parameter is always accompanied by the tension, so in practice, it is the same as taking the tension to zero. Note that in the GCA there is a single Virasoro leftover, which indicates the chiral nature of the null string. This leftover Virasoro cannot be identified with either the left, or the right moving Virasoro of the original tensionful string, it is a non-chiral linear combination of both. This symmetry algebra is also the first connection between the null string and the ambitwistor string since the algebra of constraints of the bosonic ambitwistor string is precisely a GCA.

Geometrically the ultra-relativistic limit of the string is obtained by scaling $\tau\rightarrow\epsilon\tau$ and taking $\epsilon\rightarrow0$. Thus instead of the two Virasoro algebras which locally generate the diffeomorphisms in time and space we have
\begin{equation}
 \label{eq:vec-realization}
 L_n=\im e^{\im n\sigma}(\d_\sigma+\im n\tau\d_\tau),\,\,\,\,M_n=\im
 e^{\im n\sigma}\d_\tau
\end{equation}

The relativistic algebra of the string has also another limit, the
non-relativistic one. In this case the contracted algebra is
isomorphic to the GCA of the ultra-relativistic limit. This is a
coincidence that happens in two spacetime dimensions and it is not
true in higher dimensions. This limit corresponds to contracting the
spatial dimension by $\sigma\rightarrow\epsilon\sigma$ and taking
$\epsilon\rightarrow0$. Locally, there's no difference between taking
this limit of the ultra-relativistic one, so it is no surprise that
the contracted algebras coincide. In terms of the null string, this
contraction can be thought of as corresponding to picking the dual
transverse gauge, since the roles of $\tau$ and $\sigma$ are exchanged
compared to the ultra-relativistic limit. Here this limit won't play a
role so we won't discuss it further, for more details
see~\cite{Bagchi:2009pe,Bagchi:2013bga,Bagchi:2015nca}. It is also known that the GCA2 algebra is isomorphic to the
BMS3 algebra, \cite{Campoleoni:2016vsh,Barnich:2014kra,Barnich:2014cwa,Barnich:2010eb}. Asymptotic symmetry algebras also appears in the Ambitwistor string when written in coordinates appropriate for a description as having as target space $\mathscr{I}$, the null boundary of Minkowski spacetime \cite{Geyer:2014lca}. Furthermore, there are four dimensional models which have $\mathscr{I}$ as their target space from the very beginning \cite{Adamo:2014yya,Adamo:2015fwa}. There seems to be an interesting connection between the asymptotic geometry and these models which we leave for future work.

\subsection{Quantization and spinning string}

The quantization of the null string is done in the standard way by
replacing Poisson brackets by commutators $\{\,,\}_{PB} \to
-i[\,,]$. But this only defines the quantum algebra, which now allows
certain central terms in the constraint algebra \eqref{eq:algebra}. In
order to define the quantum theory it is also necessary to pick a
representation space for this algebra, that is, a choice of vacuum, or
equivalently an ordering prescription for the operators.
Here, contrary to the tensionful string, there are two inequivalent
consistent choices. The first one was discovered
first~\cite{Lizzi:1986nv}. 
It prescribes that the $p_n$ modes should be on the right of the $y_n$ modes, and accordingly, defines the vacuum as the state annihilated by all the $p_n$ modes
\begin{equation}\label{eq:p_vacuum}
p_n|0\rangle=0\,, \;\forall\; n\in \mathbb{Z}\,.
\end{equation}
This is the so-called Weyl ordering. With this choice of vacuum, it
can be shown that the quantized GCA algebra \eqref{eq:algebra} does not have a central extension. In particular, the Virasoro subalgebra has zero central charge, implying the fact that Weyl-ordered null strings have no critical dimension~\cite{Lizzi:1986nv}. It was proposed in \cite{Lizzi:1986nv,Gamboa:1989zc} that the spectrum of this theory consists of a \textit{mass continuum} of free-particles of arbitrary integer higher spins.


This theory is not the ambitwistor string, which has a finite spectrum
of massless particles and critical dimension $26$ in the bosonic case.
But there exists another quantization prescription, which corresponds
to pulling to the right all the \textit{positive} modes of both $y_n$
and $p_n$, and shifting the negative modes to the left. This is known as the
normal ordering prescription. The vacuum is then defined as the state
annihilated by the positive modes
\begin{equation}\label{eq:cft_vacuum}
 p_n|0\rangle=y_n|0\rangle=0 \;\text{iff}\; n>0.
\end{equation}

Given this ambiguity in the quantization, which vacuum is the most natural from the $T\rightarrow0$ limit of the usual
string? In the literature, it has been observed
that~\eqref{eq:p_vacuum} is the natural limit of the string theory
vacuum. In the usual string the modes can be defined as
\begin{align}
\label{eq:string-modes-1}
 \alpha_n &= \frac{1}{2\sqrt{T}}p_n -\im n\sqrt{T}y_n\\
 \tilde{\alpha}_n&=\frac{1}{2\sqrt{T}}p_{-n}-\im n\sqrt{T}y_{-n}.
\end{align}
In the tensionless limit the vacuum conditions
$\alpha_n|0\rangle=\tilde\alpha_n|0\rangle=0$ for all $\;\forall\;n>0$
reduce to $p_n|0\rangle=0$ for all $n\in \mathbb{Z}$. But this limit
is singular and it is not at all clear that the modes $\alpha$ and
$\tilde\alpha$ are well-defined by themselves in the limit.
Nevertheless, the spectrum in the Weyl quantization is closer to the
high-energy string, since it contains an infinite tower of higher spin
states.

The normal ordering prescription is closer to what we would get from a
chiral CFT. If we imagine that the modes $y_n$ and $p_n$ are packaged in chiral fields, the
usual radial quantization would give the vacuum~\eqref{eq:cft_vacuum}.
With this prescription, it can be shown, using a variety of
methods~\cite{Gamboa:1989zc,Gamboa:1989px,Hwang:1998gs}, that the CGA
has a central extension given by a central charge in the Virasoro
subalgebra. This gives a critical dimension of $26$ for the null
string, the same as for the usual string and for the ambitwistor
string. Moreover, in this prescription the spectrum is truncated to a
finite number of massless modes, which turn out to be the same states
as the ones present in the ambitwistor string. The analysis that we
present here then sheds some light on why the bosonic (or heterotic) ambitwistor strings do not work at the quantum level.

In terms of the spectrum, the key difference between these two
quantizations lies in the definition of the quantum operator
$L_0=\sum_n y_n\cdot p_{-n}$.
In the Weyl ordering, the normal ordering constant turns out to be zero, but in the normal ordering its value is $-2$. Recall that $L_0$, being the zero mode of $\dot X\cdot X'$, generates global translations around $\sigma$, so in the latter quantization scheme the spectrum is truncated to operators with angular momentum $-2$. This is in essence the same phenomenon that arises in the bosonic ambitwistor string, where the physical states are required to have conformal weight 2 when the ghosts are stripped off. We omit the proofs of these assertions since they can be easily found in the literature, see \cite{Gamboa:1989zc} for instance. We also sketch the calculation for the spinning string below.

The three physical states are $p_{-1}^\mu p_{-1}^\nu|0\rangle$, $(p_{-1}^\mu y_{-1}^\nu -y_{-1}^\mu p_{-1}^\nu)|0\rangle$ and $p_{-1}^\mu p_{-1}^\mu|0\rangle$, corresponding to a graviton, a two-form and a dilaton, as the bosonic ambitwistor string. Note that using the usual hermitian conjugation, it is evident that these states have zero norm, which points to some inconsistency of the quantum theory. If the bosonic ambitwistor string and the null string are, as we argue below, the same then this inconsistency is to be expected, since from \cite{Adamo:2014wea} we know that the bosonic ambitwistor string is not target-space diffeomorphism invariant at the quantum level. 

From the same work, we know that the spinning version of the
Ambitwistor string with two real fermions on the worldsheet is target
space diffeomorphism invariant. This shows that it is better then to
study a supersymmetric version of the null string, which should have
better quantum behaviour. 
In \cite{Gamboa:1989zc,BarcelosNeto:1989gs}, the authors also studied
a spinning version of the null string, and showed how it appears
classically from the $T\rightarrow0$ limit of the $\mathcal{N}=(1,1)$
string. The ungauge-fixed LST action \cite{Lindstrom:1990ar} is 
\begin{equation}
  \label{eq:null-sp-str}
  S=  \int d^2\sigma\left(
    (V^\alpha \partial _\alpha X^\mu+i\Psi^{a\mu} \chi^a )
    (V^\beta \partial _\beta X^\nu+i\Psi^{b\nu} \chi^b )
    +i\Psi^{a,\mu}V^\alpha\partial_\alpha\Psi^a_\mu\right).
\end{equation}
where the fermions $\Psi^{a\mu}$ transform as densities of weight
$-1/4$. The index $a$ takes values in $\{1,2\}$ reflecting
the two supersymmetries from the tensionful $\mathcal{N}=(1,1)$
string. 
%
In the transverse gauge, the bosonic mode expansion is the same as \eqref{eq:mode-exp}, while the fermions (in the NS sector) are expanded as
\begin{equation}
 \label{eq:fermion-modes}
 \Psi^{a\mu}(\sigma)=\sum_{r\in\mathbb{Z}+\frac{1}{2}}\psi_{r}^{a\mu}e^{\im r\sigma}.
\end{equation}
The constraint algebra of these models was studied thoroughly in \cite{Gamboa:1989px,Gamboa:1989zc,Saltsidis:1995qr}. The constraints are
\begin{equation}
\label{eq:spinning-cons}
\begin{aligned} 
P^2&=0\\ 
P\cdot X' +\tfrac{\im}{2}\textstyle{\sum_{a=1,2}}\Psi^a\cdot\Psi'^{a}&=0\\
\Psi^a\cdot P&=0 \qquad(a=1,2)
\end{aligned}
\end{equation}
which are nearly the same as the ambitwistor string. The difference is, once more, that in the ambitwistor the primes in the second constraint are replaced by holomorphic derivatives. Nevertheless the modes algebra is the same for both sets of constraints
\begin{equation}
\label{eq:spinning-modes-al}
\begin{aligned}
&[L_m,M_n]=(m-n)M_{m+n}\\
&[L_m,L_n]=(m-n)L_{m+n}\\
&[L_m,G_r^a]=\left(\frac{m}{2}-r\right)G^a_{m+r}\\
&[G_r^a,G_s^b]_+=\delta^{ab}M_{r+s}
\end{aligned}
\end{equation}
where $G_n^a$ are the modes of $\Psi^a\cdot P$.

The quantization ambiguity is still present in this case. Choosing the
normal ordering prescription for the bosonic and fermionic modes
implies a non-zero normal ordering constant in the constraint
$P\cdot X' +\frac{\im}{2}\Psi^a\cdot\Psi'^a$, which in turn gives a
critical dimension 10 in the case of two fermions.\footnote{Mixed
  boundary conditions turn out to be inconsistent~\cite{Gamboa:1989zc}.} 
The spectrum in the NS vacuum is given by $p_{-1}|0\rangle$ and
$\psi^a_{-1/2}\psi^b_{-1/2}|0\rangle$, but after a GSO projection it
is truncated to only $\psi^1_{-1/2}\psi^2_{-1/2}|0\rangle$. This
state, together with the NS-R and R-R sectors, reproduces the spectrum
of type II A/B supergravity. As expected these results are the same as
the type II ambitwistor string. What seems to have prevented the
authors of \cite{Gamboa:1989zc} from calculating
scattering amplitudes was a lack of knowledge of the vertex operators
in their formalism. Here is where the ambitwistor string shines. Since
it is written as a CFT, constructing vertex operators is trivial, and
scattering amplitudes can be easily calculated via OPE's. Models with
more supersymmetry were studied in \cite{Saltsidis:1995qr}, where it
is seen that the $\mathcal{N}=4$ model has the same critical dimension
as found by Ohmori in~\cite{Ohmori:2015sha} for the ambitwistor string
with extra SUSY. These connect to the zoo of CHY models \cite{Cachazo:2014xea,Casali:2015vta}. All of this is overwhelming evidence that the
ambitwistor string and the null string are the same physical system,
provided the correct quantization scheme is chosen for the null
string. In the next section we show that how the ambitwistor string
can be obtained from the null string and we study some of its
peculiarities.

\section{The ambitwistor string}

We turn now to the ambitwistor string. Our objective here is to describe how it fits into the framework of the null string as introduced above, and thus, clarify its relation to the usual tensionful string.


So far, we have avoided discussing a crucial point concerning boundary conditions. Starting from a tensionful string with the standard periodicity $\sigma\simeq\sigma+2\pi$, we took the null limit and kept this periodicity throughout the process. However, the null string worldsheet has a degenerate metric,  so there are no canonical time-like and space-like directions. A time-like direction is needed to perform the usual canonical quantization, and to identify a worldsheet Hamiltonian. It is also not guaranteed from the start that space-like periodicity is not going to clash with the choice of time-like direction. 
The Schild gauge essentially assumes that the time is given by the direction of $V^\alpha$. Given the natural pairing $Z_\alpha W^\alpha=Z_0 W^0 + Z_1 W^1$,  a natural choice for the periodic coordinate is in the direction of a $U_\alpha$ in the kernel of $V^\alpha$. But this solution is not forced upon us and other choices are possible. Indeed the ambitwistor string is one. We will see that the choice $\mu=1$, and the requirement of periodicity along the $1$ (or $\sigma$) direction will be possible only in the gauge $\lambda=0$.

\subsection{The ambitwistor string action}

The ambitwistor string was originally formulated as a holomorphic CFT in Euclidean signature. In order to make contact with the null string we Wick rotate back to Lorentzian signature and write the original action as
\begin{equation}\label{eq:O-action}
 S_{A}=\int d^2\sigma\; P\cdot\d_-X
\end{equation}
where $\sigma^{\pm}=\tau\pm\sigma$ and $\partial_{\pm}=\frac12(\partial_\tau\pm\partial_\sigma)$. This action is supplemented by the constraints
\begin{equation}
 T_{++}=P\cdot\d_+X=0,\;\;\; P^2=0.
\end{equation}
The null string first order action is
\begin{equation}
  S=\int d^2\sigma P\cdot\dot{X} - \mu P\cdot X' - \lambda P^2.
\end{equation}
Taking $\mu=1$ and $\lambda=0$ we get the ambitwistor string, but with slightly different constraints
\begin{equation}
 P\cdot X'=0,\quad P^2=0.
\end{equation}
However, using the ambitwistor equation of motion $\d_-X=0$ 
\begin{equation}
P\cdot X'=P\cdot(\d_--\d_+)X=P\cdot\d_+X
\end{equation}
and we used the ambitwistor string equation of motion $\d_-X=P$, together with the $P^2=0$ constraint. The ambitwistor string constraints and the null string constraints are then the same on-shell, therefore the strings are classically equivalent.

Although a simple fact, the importance of the Ambitwistor gauge can't
be overlooked. Choosing the gauge $\mu=1$ effectively picks a
background light-cone structure for the worldsheet. In the Wick rotated framework, this is equivalent to a choice of background complex structure, and a choice of a metric.\footnote{More precisely, a conformal class of metrics.} This is an emergent metric coming purely from the gauge-fixing, the null string has no such non-degenerate metric. The components of the null string stress tensor~\eqref{eq:stress_energy} are
\begin{align}\nonumber
T_1{}^0&=P\cdot X'\\
T_0{}^0&=-T_1{}^1=2\lambda P^2 +\mu P\cdot X'\\
T_0{}^1&=-\mu(4\lambda P^2+\mu P\cdot X')\nonumber
\end{align}
In the Ambitwistor gauge ($\lambda=0,\;\mu=1$) and on-shell, the
stress tensor of this emergent complex structure coincides with the
null string stress-energy tensor. This significantly simplifies
finding the moduli for the gauge-fixed null string, as it now
coincides with finding the moduli for the effective complex structure.
It also enables us to use standard CFT methods to study the null
string and compute correlation functions. We expect that a similar
mechanism is at work in the other four dimensional twistor strings
\cite{Witten:2003nn,Berkovits:2004hg,Skinner:2013xp,Geyer:2014fka}, and that they all fit in the framework of null strings. In this regard we mention that in \cite{Siegel:2004dj,Bandos:2014lja}, a relationship between the four dimensional twistor strings and the null string was already noted.

A more geometrical picture is given by the second-order LST action~(\ref{eq:S-V}).
The Ambitwistor gauge is singular in the second-order framework since the action is proportional to $\lambda^{-1}$. In order to study it, we go to an intermediary gauge similar to what is called the HSZ gauge in Siegel's paper~\cite{Siegel:2015axg,Hohm:2013jaa}, although there it was used in the tensionful string. It is given by
\begin{equation}
  \label{eq:V-ambi}
  V^\alpha=\frac1{2\sqrt{\lambda}}(1,-1).
\end{equation}
In terms of the Lagrange multipliers of the Hamiltonian action this choice corresponds to $\mu=1$ and $\lambda$ unfixed. While the gauge fixing condition $\lambda=0$ is perfectly fine in the first order action, it is clearly singular in the second order action
The (partially) gauge-fixed action reads
\begin{equation}
  \label{eq:S-lightcone}
  S=-2\int d^2\xi \frac{1}{\lambda}(\partial_-X)^2 \,.
\end{equation}
Since $\lambda$ has the correct dimension to be a worldsheet
loop-counting parameter (like $\hbar$ or $\alpha'$), the limit
$\lambda\to0$ appears as a classical limit, and the action is expected
to localize on the extremal field configurations, i.e. the classical solutions. In other words, the string partition function
  $\int \mathcal{D}X e^{-\frac1\lambda S[X]}$
is independent of $\lambda$, which we can be taken to zero.

\subsection{Residual gauge symmetry}
\label{sec:resid-gauge-symm}
We follow here an analysis of~\cite{Isberg:1993av}, which we perform it in the ambitwistor gauge.
In the tensionful string, the residual symmetry that remains after gauge fixing are the conformal transformations of $z$ and $\bar z$, here the residual symmetries form the Galilean Conformal transformations, generated by the vector field $\epsilon^\alpha$ or $\epsilon^\pm=\epsilon^0\pm\epsilon^1$ with 
\begin{equation}
\label{eq:epsilon-def}
 (\epsilon^+,\epsilon^-)=(\sigm f'(\sigp)+g(\sigp),f(\sigp))
\end{equation}
where, in this section only, prime is the ordinary derivative, not the $\sigma$ derivative. 
By comparison, conformal transformation are generated by vector fields
  $(\epsilon^+,\epsilon^-)=(f(\sigp),g(\sigm))$.
On a function $F(\sigm,\sigp)$, the vector field
(\ref{eq:epsilon-def}) generates the following transformations;
\begin{equation}
  \label{eq:delta-F}
  \delta F=\left[(\sigm f'(\sigp)\partial_-+f(\sigp)\partial_+)+g(\sigp)\partial_-\right]F
\end{equation}
from which we define
\begin{equation}
  \label{eq:LM-def}
  L(f)=\sigp f'(\sigp)\partial_-+f(\sigp)\partial_+,\quad M(g)=g(\sigp)\partial_-
\end{equation}
whose algebra is obtained by decomposing $f$ and $g$ in modes 
\begin{equation}
f=\sum_{n} f_n e^{in\sigp}\;\;\;g=\sum_n g_ne^{in\sigp} 
\end{equation}
such that
\begin{equation}
  \label{eq:modes-LM}
L(f)=-i\sum_n f_nL_n\,,\quad M(g)=-i\sum_n g_n M_n
\end{equation}
where

\begin{align}
  L_n
=i e^{in\sigp}
  (\partial_++in
  \sigm \partial_-)\,,\quad \label{eq:LM_npm}
  M_n
=i e^{in\sigp}\partial_-\,.
\end{align}

They satisfy the following commutation relations;
\begin{equation}
  \label{eq:gca-classical}
  [L_n,L_m]=(n-m)L_{n+m},\quad [L_n,M_m]=(n-m)M_{n+m},\quad [M_n,,M_m]=0\,.
\end{equation}
It seems that in the Ambitwistor gauge we have lost the periodicity of
the original null string. In the next section we shall show that this is merely an artifact of this gauge choice.

\subsection{Equations of motion in the second order form}
\label{sec:equations-motion-2nd}

Using the second order action in chiral gauge (\ref{eq:S-lightcone}), the equations of motion are 
\begin{equation}
  \label{eq:eom-lightlike}
  \partial_-\partial_-X=0,
\end{equation}
subject to the constraints 
  \begin{align}
  \label{eq:Psq-SO}
    ( \partial_-X)^2=0\\ 
  \label{eq:PdX-SO}
\partial_-X \cdot\partial_\sigma X=0
  \end{align}
where the constraint (\ref{eq:Psq-SO}) is the EOM for $V^\alpha$, just
like in the standard string, where the EOM for the auxiliary metric condition gives the Virasoro constraints $T_{\alpha\beta}=0$.
The EOM are solved by 
\begin{equation}
  \label{eq:sol-eom}
  X(\sigm,\sigp)=Y(\sigp)+\sigm P(\sigp)
\end{equation}
with mode expansion
%
%

\begin{equation}
 Y(\sigp)=\frac1{\sqrt{2\pi}}\sum_{n}y_n e^{-\im n\sigp},\quad K(\sigp)=\frac1{\sqrt{2\pi}}\sum_{n}k_n e^{-\im n\sigp}.
\end{equation}
It seems that the solution \eqref{eq:sol-eom} breaks the string's periodicity $\sigma_{\pm}\simeq\sigma_{\pm} \pm 2\pi$. In order to restore it we impose the periodicity conditions
\begin{align}
 &K(\sigp + 2\pi)=K(\sigp)\\
 &Y(\sigp+2\pi)\simeq Y(\sigp) +2\pi K(\sigp),
\end{align}
or in terms of the modes $y_n\simeq y_n +2\pi k_n$. While this seems to be a strange condition to impose, it is consistent with the Poisson brackets
\begin{align}
\{y_n,y_m \}&\simeq \{y_n + 2\pi k_n,y_m + 2\pi k_m\}=0\\
\{y_n,k_m\}&\simeq\{y_n + 2\pi k_n,k_m\}=\delta_{m+n,0}.
\end{align}
One might worry about the constraint algebra, and indeed at first
sight it seems that there will be problems in defining the modes for
$\d_-X\cdot X'$. Ultimately, these issues stem from a tension between
the periodicity in the gauge $\mu=1$ and the singularity of the gauge
$\lambda=0$ that we are trying to impose in the second-order action. Note that the canonical momentum depends on $\lambda$ 
\begin{equation}
 P=\frac{1}{\lambda}\d_-X=\frac{1}{\lambda}K
\end{equation}
so, in order to have a finite canonical momenta when $\lambda\rightarrow0$ we need to rescale the modes of $K$ by $\lambda$. This rescaling leaves the first constraint \eqref{eq:Psq} invariant, but does affect the second constraint \eqref{eq:PdX}
\begin{equation}
 \frac{1}{\lambda}\d_-X\cdot X'=P\cdot (Y' - K - \sigma K')=P\cdot (Y' -\lambda P - \lambda\sigma P')\rightarrow P\cdot Y'
\end{equation}
where all the constraints are evaluated at $\tau=0$ for simplicity. We
see that in the limit we recover the first-order constraints, as
expected, and that the periodicity condition on the modes $y_n$
disappears. This shows that to work with the more geometric
second-order Lagrangian we should be very careful in order not to run
into inconsistencies. 

\subsection{Equations of motion in the first order form}
\label{sec:equations-motion}

Using the first order action in the $\mu=1$ gauge, with $\lambda$ yet
unfixed, the EOM for $P$ and $X$ read
\begin{equation}
\label{eq:eom-lightlike2}
\partial_-X=2\lambda P\,,\qquad\partial_-P=0
\end{equation}
still subject to the constraints
  \begin{align}
  \label{eq:Psq}
    P^2=0\\ 
  \label{eq:PdX}
P\cdot\partial_\sigma X=0
  \end{align}
The EOM are solved by 
\begin{equation}
  \label{eq:sol-eom2}
  P=P(\sigm)\,,\qquad X=Y(\sigp)+2\lambda \sigm P(\sigp)
\end{equation}
with mode expansion
\begin{equation}
 Y(\sigp)=\frac1{\sqrt{2\pi}}\sum_{n}y_n e^{-\im n\sigp},\quad 
 P(\sigp)=\frac1{\sqrt{2\pi}}\sum_{n}p_n e^{-\im n\sigp}.
\end{equation}
It seems a priori that the solution \eqref{eq:sol-eom2} breaks the periodicity
$\sigma_{\pm}\simeq\sigma_{\pm} \pm 2\pi$. 
But setting $\lambda=0$, the $\sigma$-periodicity is restored
and the mode expansion of the $PX'$ constraint simplifies to the
following;
\begin{equation}
\label{eq:constr-mode-PdX}
 P\cdot X'\big|_{\lambda=0} =
 P(\sigp)\cdot \partial_\sigma Y(\sigp)=-\frac{\im}{2\pi}\sum_{n,m}n y_{-m}\cdot p_{n+m} e^{-\im n\sigp}=\frac{1}{2\pi}
\sum L_n e^{-in \sigp
}\end{equation}
($P^2$ does not change). 

We now give a somewhat heuristic argument to support the idea that the
singular gauge $\lambda=0$ really allows to actually gauge the
stress-tensor of a chiral CFT and not just the $P\cdot X'$
constraint. 
For arbitrary $\lambda$, it is not hard to see that
\begin{equation}
  P\cdot \partial_+ X \propto \sum_n (L_n-in \lambda \sigm M_n)e^{-in\sigp}\,.
\end{equation}
with $L_n$ and $M_n$ defined as in eq.~\eqref{eq:modes-cons}. 
Hence, at $\lambda=0$, we see that the $L_n$'s, which are the modes of
the null string constraints really do coincide with the modes of the
stress-tensor. 

On the other hand, it is evident, from looking at the Lie algebra
representation of the $L$ and $M$'s in  eq.~(\ref{eq:LM_npm}), that
the combination $L_n'\equiv(L_n-in \sigm M_n)= -\im e^{-in\sigp}\partial_+$
are the generators of conformal transformations on $\sigm$. Moreover,
the $L_n'$ generator obey the same algebra with the $M_n$'s, so this
is an automorphism of the GCA algebra.\footnote{We are grateful to
  Balgoje Oblak for a discussion on this point.} So it seems that the
singular gauge choice $\lambda=0$ automatically enforces that the
constraint algebra of the tensionless string do become that of the
holomorphic stress-tensor of a chiral CFT.

\section{Relation to tensionful strings}
\label{sec:relat-tens-strings}

\subsection{The bosonic string}
\label{sec:bos_string}
At the classical level, the relationship between the ambitwistor
string and the tensionful string is straightforward. As we have shown
above, the classical ambitwistor string is nothing else than the
tensionless limit of the usual string in an unusual
gauge. It is at the quantum level that
the relationship becomes interesting. This is due to the non-standard
(from the tensionful string point of view) choice of vacuum for the
ambitwistor string. Recall that the ambitwistor vacuum is defined as
the state annihilated by the positive modes of $P$ and $X$, see~\eqref{eq:cft_vacuum}.
To make contact with the usual string, recall the definition of the modes
\begin{equation}
\begin{aligned}
 \alpha_n &= \frac{1}{2\sqrt{T}}p_n -\im n\sqrt{T}y_n\\
 \tilde{\alpha}_n&=\frac{1}{2\sqrt{T}}p_{-n}-\im n\sqrt{T}y_{-n}
\end{aligned}
\end{equation}

which inherit the commutation relations
\begin{equation}
 [\alpha_n,\alpha_m]=[\tilde{\alpha}_n,\tilde{\alpha}_m]=n\delta_{m+n,0}
\end{equation}
from the canonical ones $[y_n,p_m]=i\delta_{n+m,0}$. In terms of these modes, the ambitwistor vacuum obeys
\begin{equation}\label{eq:alt_quant}
 \alpha_n\vacxp=\tilde\alpha_{-n}\vacxp=0\,,\quad\forall\,n>0\,,
\end{equation}
in stark contrast to the string theory vacuum, defined by
\begin{equation}
 \alpha_n|0\rangle_S=\tilde\alpha_n|0\rangle_S=0\,,\quad\forall\,n>0.
\end{equation}
The alternative choice of vacuum (\ref{eq:alt_quant}) for the
tensionful string was briefly considered in~\cite{Hwang:1998gs}. The
authors noted that, although the vacuum is not BRST-invariant there
are physical massless states. Unfortunately, these have negative
norm.
Interestingly this choice of vacuum changes the normal ordering
constant of the $\tilde{\mathcal{L}}_0$ mode from $-1$ to
$1$. This is due to the fact
that a consistent BRST quantization of this vacuum requires that
$\tilde{c}_{-m}|0\rangle_A=0$ for all $n>0$, in contrast to the
right-handed condition $c_m|0\rangle_A=0$ for $n>0$. The
the Virasoro generators are
\begin{align}
\mathcal{L}_m &= \frac{1}{2}\sum_n: \alpha_{m-n}\cdot\alpha_{n}:+\sum_n(2m-n):b_nc_{m-n}: +a\delta_{m,0}\\
\tilde{\mathcal{L}}_m &= \frac{1}{2}\sum_n:\tilde{\alpha}_{m-n}\cdot\tilde{\alpha}_{n}:+\sum_n(2m-n):\tilde{b}_n\tilde{c}_{m-n}: +\tilde{a}\delta_{m,0}
\end{align}
The normal ordering constant is obtained by standard methods which we
reproduce here to highlight the difference with the usual
string. Starting from $(\mathcal{L}_0-a)\vacxp=0$ and
$(\mathcal{\tilde L}_0-\tilde a)\vacxp=0$, we use
\begin{align}
  2\mathcal{L}_0|0\rangle_A  = (\mL_1\mL_{-1}
  -\mL_{-1}\mL_1)|0\rangle_A&=\mL_1\mL_{-1}|0\rangle_A
=-(2b_0c_1)(b_{-1}c_0)|0\rangle_A\\ \nonumber
  &=-|0\rangle_A
\\
  2\tilde{\mathcal{L}}_0|0\rangle_A = (\tilde{\mL}_1\tilde{\mL}_{-1} -\tilde{\mL}_{-1}\tilde{\mL}_1)|0\rangle_A&=-\tilde{\mL}_{-1}\tilde{\mL}_1|0\rangle_A=(2\tilde{b}_0\tilde{c}_{-1})(\tilde{b}_{1}\tilde{c}_0)|0\rangle_A\\ \nonumber
  &=+|0\rangle_A\,.
\end{align}
So the normal ordering ambiguity of the operator
$\mathcal{L}_0+\tilde{\mathcal{L}}_0$, which contains the mass-shell
condition, is ``transferred'' to the angular momentum operator
$\mathcal{L}_0-\tilde{\mathcal{L}}_0$. Physical states are then
restricted to those with angular momentum $2$ and the spectrum is
truncated. Contrary to the null string the spectrum is not all
massless, there are two massive states with the same, but opposite
mass proportional to the tension $T$.\footnote{This was also noted in
  \cite{Hwang:1998gs,Huang:2016bdd}.} Consistency of the quantization
procedure also requires that the vacuum obeys
\begin{equation}
 \mathcal{L}_n|0\rangle_A=\tilde{\mathcal{L}}_{-n}|0\rangle_A=0\,,\quad\forall\,n>0,
\end{equation}
this effectively exchanges the roles of translations and special conformal transformations in the left-handed part of the conformal algebra. That is, we're picking different representation spaces for the right-handed Virasoro and for the left-handed Virasoro.

Going one step further we can define the modes
\begin{align}\label{eq:mode_comb}
 L_n=(\mathcal{L}_n-\tilde{\mathcal{L}}_{-n})&=-\im\sum kp_{n-k}\cdot
                                               y_k -2\delta_{n,0}\\
 M_n=T(\mathcal{L}_n+\tilde{\mathcal{L}}_{-n})&= \frac{1}{2}\sum_k\frac{1}{2}p_{n-k}\cdot p_k+2T^2(k-n)ky_{n-k}\cdot y_k.
\end{align}
In the $T\to0$ limit they coincide with the constraint modes of the
Ambitwistor string, including the normal ordering constant. This change of sign was also noted in \cite{Siegel:2015axg,Huang:2016bdd}, but there it
was attributed to some kind of change in boundary conditions for the bosonic string $X$ field. As explained above our point of view is that this is the result of a twisted quantization schemes for the string. 

We further note that the ambitwistor vacuum $|0\rangle_A$ and the
string theory vacuum $|0\rangle_S$ are not related by a Bogolioubov
transformation, they give inequivalent quantizations of the classical
string. That is, these vacua live in different, unitarily
inequivalent, Hilbert spaces. This implies that the quantum
Ambitwistor string is \textit{not} a subsector of the usual string in
any natural way. This also explains why the Ambitwistor string does
not come from the $T\rightarrow\infty$ limit of the string (field
theory or $\alpha'\to0$ limit), since it is in fact the null string in
disguise. One last thing to note is that the combination of modes
\eqref{eq:mode_comb} corresponds to the well-defined modes in the ultrarelativistic limit of the Virasoro algebra, which is equivalent to the tensionless limit of the algebra~\cite{Bagchi:2013bga,Bagchi:2015nca}. 

To close this section we mention that above alternative quantization
of the string can be achieved by a slight modification of the
Ambitwistor string. While still using the chiral action \eqref{eq:O-action}, we modify the constraint
\begin{equation}\label{eq:new_const}
\mathcal{H}=P^2+(\partial X)^2.
\end{equation}
Using the free OPEs, it is easy to see that the algebra of constraints is modified to
\begin{align}\nonumber
\mathcal{T}(z)\mathcal{T}(w)&\simeq \frac{2\mathcal{T}}{(z-w)^2} + \frac{\partial\mathcal{T}}{(z-w)}\\\nonumber
\mathcal{T}(z)\mathcal{H}(w)&\simeq \frac{2\mathcal{H}}{(z-w)^2} + \frac{\partial\mathcal{H}}{(z-w)}\\
\mathcal{H}(z)\mathcal{H}(w)&\simeq -\frac{8\mathcal{T}}{(z-w)^2} + -\frac{4\partial\mathcal{T}}{(z-w)}
\end{align}
where $\mathcal{T}=-P\cdot \partial X$ is the stress-energy tensor. The massless sector consists of the BRST closed vertex operators
\begin{align}\nonumber
U &= c\tilde{c} \epsilon_{\mu\nu}(P^\mu P^\nu - \partial X^\mu\partial X^\nu)e^{\im k\cdot X}\\
B &= c\tilde{c} \epsilon_{\mu\nu}P^{[\mu}\partial X^{\nu]}e^{\im k\cdot X}\\ \nonumber
D &= c\tilde{c} (P^2-(\partial X)^2)e^{\im k\cdot X}
\end{align}
corresponding to a graviton, b-field and dilaton, respectively. The fields $c$ and $\tilde{c}$ are the ghosts for the constraints, see \cite{Mason:2013sva}. Contrary to the original bosonic Ambitwistor string, the cohomology now allows for massive states. There are two massive spin two states given by
\begin{equation}
V=c\tilde{c} \epsilon_{\mu\nu}(P^\mu P^\nu + \partial X^\mu\partial X^\nu \pm P^{(\mu}\partial X^{\nu)})e^{\im k\cdot X}
\end{equation}
with masses $k^2=\mp 4$ in our conventions. Besides being a realization of the alternative quantization of the string, this is very reminiscent of the HSZ theory \cite{Hohm:2013jaa,Huang:2016bdd}. We believe that the CFT above gives a realization of that theory and might help in calculating higher-point amplitudes. The tension might be restored by dimensional analysis, and in the $T\rightarrow0$ limit the two massive spin two states become massless and indistinguishable from the graviton vertex operator. In this limit the model reduces to the original Ambitwistor string. 

\subsection{The spinning string}
\label{sec:spin_string}

The bosonic string in the alternative quantization scheme
\eqref{eq:alt_quant} has negative normed states which are the same as
the ones found in the null string. The next step is then to add
supersymmetry and see if the resulting model has only positive normed
states in analogy with the null string. We will see that this is
indeed the case, though the resulting quantization procedure looks
strange from the original worldsheet perspective.\footnote{Instead of
  the requirement that right movers are holomorphic around the origin
  and left movers antiholomorphic around the origin, the twisted
  quantization requires right movers to be holomorphic around the
  origin and left movers to be antiholomorphic around infinity.}

Classically the supersymmetry is the same as the usual $(1,1)$ string, that is, the fermionic generators are
\begin{equation} 
\mathcal{G}_r=\sum\alpha_n\cdot\psi_{r-n}\,,\quad\tilde{\mathcal{G}}_r=\sum\tilde{\alpha}_n\cdot\tilde{\psi}_{r-n}.
\end{equation}
To have a consistent BRST quantization their action on the vacuum must
be different form the usual string, in particular
$\tilde{\psi}_r|0\rangle_A=0$ for all $r>-\frac{1}{2}$. This has a
cascading effect for the left-handed fields. The bosonic ghost vacuum
now obeys $\tilde{\gamma}_{-n}|0\rangle_A=0$ for $n>0$ which
contributes  $+\frac{1}{2}$ to the normal ordering constant of
$\tilde{\mL}_0$. The end result is that all the left-handed fields
have their quantization flipped, that is, their positive modes now
annihilate the vacuum, and the spectrum is truncated to states
with angular momentum $1$. This is analogous to the bosonic case, but
now the GSO projected physical states are built out of the fermions,
$\psi_{-1/2}\tilde{\psi}_{1/2}|k\rangle_A$, instead of the modes of
$X$. Once gauge-invariances are taken into account\footnote{This is
  most easily done by going to lightcone gauge.} the physical states
have positive norm. This reflects the situation for the null spinning
string, but here the tension has been kept finite. It is interesting
to note that, like in the bosonic case, there are possible massive
states, of the form $\psi\psi$ and $\tilde\psi\tilde\psi$. However,
they are projected out by the GSO projection. So, even at finite $T$
the only physical states in the spectrum are massless.
    
It is easy to see that the algebra of constraints limits nicely into
the ambitwistor string constraint algebra. Define the modes
\begin{align}\label{eq:ferm_modes}
 \begin{array}{ll}
 L_n=(\mathcal{L}_n-\tilde{\mathcal{L}}_{-n})\,,\quad & M_n=T(\mathcal{L}_n+\tilde{\mathcal{L}}_{-n})\,, \\ 
 G_r=\sqrt{T}\mathcal{G}_n\,, & \tilde{G}_r=\sqrt{T}\tilde{\mathcal{G}}_{-n}\,.\\
 \end{array}
\end{align}
The bosonic part of the algebra mimics what happens in the bosonic string. The relevant fermionic anticommutators are
\begin{align}
 \{G_r,G_s\}&=\{\tilde{G}_r,\tilde{G}_s\}=M_{r+s}\\
 \{G_r,\tilde{G}_s\}&=TL_{r+s}.
\end{align}
When $T=0$, this algebra is isomorphic to that of the spinning ambitwistor string \cite{Mason:2013sva}, and also coincide with the commutators found in \cite{Gamboa:1989zc} for the worldsheet supersymmetrization of the null string. As seen in the bosonic case, the tensionless limit mixes left and right-handed generators in a non-trivial way. 
We note that this algebra does not coincide with a supersymmetrization of the GCA algebra used in \cite{Mandal:2010gx}. In our case, we are interested in the $T\rightarrow0$ limit of the algebra, which corresponds to the ultra-relativistic limit of the bosonic part of the string, while in \cite{Mandal:2010gx}, the limit studied also involves contracting the odd sector.

In the purely bosonic case the algebras obtained from these two limits
are isomorphic. In the supersymmetric case there seems to be some
freedom in how the fermionic generators are rescaled and combined in
order to get finite, non-trivial generators. Clearly, the scalings
above are the correct ones in order to reproduce the spinning
tensionless string, but it would be interesting to study if the other
possible scalings give interesting theories. This limit can also be taking at the quantum level in a similar way to the bosonic case, and the central charge is seen to still vanishes only in ten dimensions.

We close this section by noting that the construction of
\cite{Bjerrum-Bohr:2014qwa}\footnote{Some observations about
  the connection between integration by parts in string theory and the scattering
  equations were also made in \cite{Gomez:2013wza}.} appears as an hybrid formulation, somehow
close to Fairlie \& Robert's original paper~\cite{Fairlie:1972zz}. The
string integrand is computed at finite tension, but the nearest
neighbour interaction term $T^2 X'^2$ is dropped in the constraint
eq.~\eqref{eq:constraints}. It remains to be understood how this fits
exactly in the rich network of theories that were discussed here and
in the aforementioned references.

\section{Discussion}

We showed that the Ambitwistor string is only related to the usual
string at the classical level, being a way of describing the
$T\rightarrow0$ limit of the string, i.e. the null string. This also
explains why the scattering equations, a feature appearing in the high
energy limit of strings also appears in a crucial way in the
Ambitwistor string. Though a more complete treatment of the moduli
problem for the null string is left for future work. It is at the
quantum level that these theories differ in a profound way. The
quantization procedures for the two theories are inequivalent and
there is no canonical way of relating the vacuum in one theory to the
vacuum in the other. Nevertheless, one might still hope to find
$\alpha'$ corrections to the Ambitwistor string keeping in mind the
issues addressed in this paper. There are also other questions to
address in future works; like other possible null strings and their
Ambitwistor description, including the tensionless limit of heterotic
and open strings, and especially the construction of the measure on the moduli
space.


\paragraph{Acknowledgements}

We would like to thank Blagoje Oblak, Lionel Mason, Nima Doroud, David
Skinner, Bo Sundborg for interesting discussions and comments. The work of ED is
supported by EPSRC grant EP/M018911/1, the work of PT is supported by
STFC grant ST/L000385/1 and Queens' College Cambridge Postdoctoral
Research Associateship.

\newcommand*{\doi}[1]{\href{http://dx.doi.org/#1}{doi: #1}}
\bibliography{biblio}  
\bibliographystyle{JHEP} 

\end{document}